\documentclass[preprint,12pt]{elsarticle}

\usepackage{amssymb}
\usepackage{amsmath}

\usepackage{xcolor}
\usepackage{siunitx}
\usepackage{bm}
\usepackage{tabularx}

\journal{Nuclear Physics B}

\begin{document}

\begin{frontmatter}


\title{Morphological Effects on Bacterial Brownian Motion: Validation of a Chiral Two-Body Model}

\author[label1,label2,label3]{Baopi Liu}
\affiliation[label1]{organization={School of Advanced Interdisciplinary Studies, Ningxia University}, city={Zhongwei}, state={Ningxia}, postcode={755000}, country={China}}
\affiliation[label2]{organization={School of Physics, Ningxia University}, city={Yinchuan}, state={Ningxia}, postcode={750021}, country={China}}
\affiliation[label3]{organization={Complex Systems Division, Beijing Computational Science Research Center}, city={Beijing}, postcode={100193}, country={China}}
\ead{bpliu@mail.bnu.edu.cn}

\author[label4,label5]{Bowen Jin}
\affiliation[label4]{organization={Huanjiang Laboratory}, city={Shaoxing}, state={Zhejiang}, postcode={311899}, country={China}}
\affiliation[label5]{organization={State Key Laboratory of Fluid Power and Mechatronic Systems, Department of Mechanics, Zhejiang University}, city={Hangzhou}, state={Zhejiang}, postcode={310027}, country={China}}

\author[label6]{Lu Chen}
\affiliation[label6]{organization={College of Physics, Changchun Normal University}, city={Changchun}, state={Jilin}, postcode={130032}, country={China}}

\author[label7]{Ning Liu}
\affiliation[label7]{organization={School of Mathematics and Physics, Anqing Normal University}, city={Anqing}, state={Anhui}, postcode={246133}, country={China}}
\begin{abstract}
We systematically investigate how flagellar morphology governs the stability of bacterial Brownian motion, evaluating the effectiveness of a simplified chiral two-body model. This model, which effectively captures the specific bacterial morphology and significantly reduces computational cost, is used for simulating bacterial Brownian motion. Our results demonstrate that the model accurately reproduces the Brownian motion of bacteria for contour lengths $\Lambda\ge5.0$~\si{\mu m}, helix radii $0.2\le R\le 0.5$~\si{\mu m}, and pitch angles $\pi/6\le\theta\le2\pi/9$. We find that the translational and rotational velocities of bacteria depend linearly on the motor rotation rate, independent of dynamic viscosity. Increasing helix radius and contour length leads to more elongated trajectories and enhances their linearity. Furthermore, longer contour lengths improve the stability of the bacterial forward motion. Collectively, these findings demonstrate the essential role of flagella in stabilizing bacterial Brownian motion and confirm the effectiveness of the chiral two-body model for simulating this phenomenon.
\end{abstract}
\begin{graphicalabstract}
\includegraphics[width=1.00\textwidth]{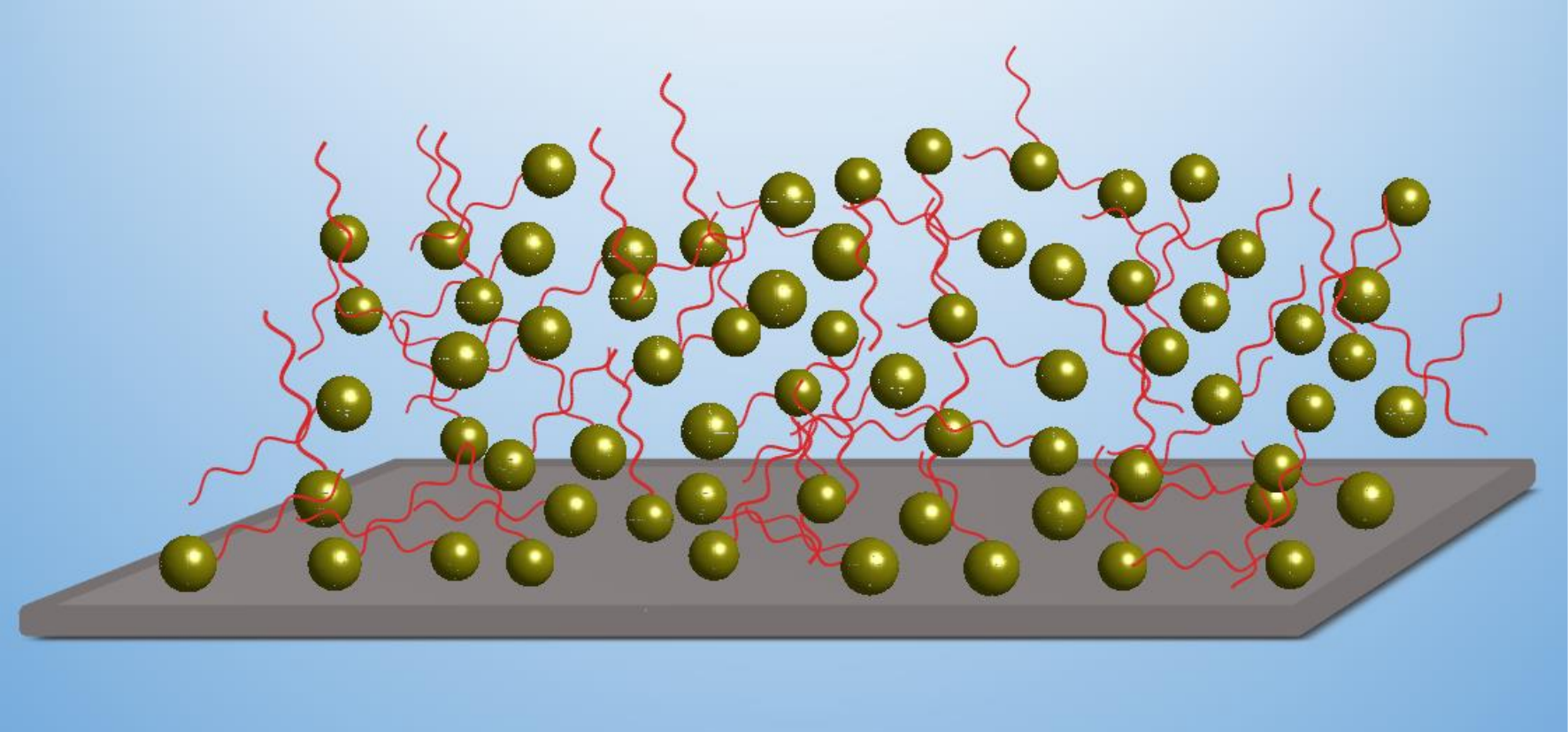}
\end{graphicalabstract}
\begin{highlights}
\item We validated the chiral two-body model for simulating bacterial Brownian motion
\item The role of bacterial flagella in stabilizing Brownian motion was studied
\item Derived analytical solutions for bacterial velocities from the chiral two-body model
\item Quantified trajectory shape, linearity, and directional stability
\end{highlights}
\end{frontmatter}
\section{Introduction}
\label{sec1}
Most bacteria swim freely in complex fluid media by generating thrust through the rotation of their flagella~\cite{Sowa2008,Lauga2016,Nord2022}. Bacteria are typically composed of a cell body, flagella, motors, and connecting structures known as hooks. The morphology of bacteria, closely related to their survival in various environments, allows them to evolve efficient adaptations that significantly influence their survival abilities and interactions with the surrounding environment~\cite{Roszak1987,Young2006,Zhang2021}. The evolution of bacterial morphology is driven by multiple biological factors, including nutrient acquisition, cell division and separation, surface adhesion, passive diffusion, active transport, and predator evasion. In competitive environments, bacteria with greater adaptability are favored and undergo continuous morphological optimization to enhance survival within ecosystems~\cite{Garcia2016,Van2017,Nguyen2019,Laventie2020,Weady2024,Liu2025B}.

Flagellated bacteria are important subjects for studying microbial motility because of their unique structures and propulsion mechanisms. Species such as \emph{Escherichia coli}~\cite{Vigeant1997,Diluzio2005,Molaei2014,Secchi2020,Lee2021} and \emph{Pseudomonas aeruginosa}~\cite{Secchi2020,Lee2021,Chang2018,Deng2020,Tian2022} are widely used in experimental, theoretical, and numerical studies due to their manipulability and relatively simple models. Although thermal noise induces Brownian motion of bacteria~\cite{Li2008,Lobaskin2008,Drescher2011}, experiments often observe directed movement~\cite{Darnton2007,Son2013,Bianchi2017,Figueroa2020,Grognot2021,Junot2022}, contrasting with the random diffusive behavior observed in active colloids~\cite{Howse2007,Jiang2010,Morin2017}. The presence of flagella not only propels the bacteria, but also stabilizes their Brownian motion. However, it remains unclear how flagellar morphology influences the stability of bacterial Brownian motion.

High-resolution simulations of bacterial motion are computationally expensive. To reduce these costs, researchers often simplify bacteria into two-body (TB) models for dynamic simulations~\cite{Dunstan2012,ZhangBo2021,Di2011}. These TB models include both achiral~\cite{Dunstan2012,ZhangBo2021} and chiral models~\cite{Di2011}, with the latter considered superior because it is obtained by integrating along the flagellar centerline and averaging the phase through resistive force theory (RFT)~\cite{Gray1955,Chwang1975,Johnson1979}, effectively captures the chirality and morphological characteristics of the flagella. Although the chiral TB model is widely used~\cite{Di2011,Dvoriashyna2021}, its validity in simulating Brownian motion still requires further exploration.

This paper has two main objectives: first, to study the influence of flagellar morphology on the stability of bacterial Brownian motion; and second, to validate the effectiveness of the chiral TB model in simulating such motion. To achieve these, we simulate the bacterial Brownian motion using the chiral TB model~\cite{Di2011}, RFT~\cite{Gray1955,Chwang1975}, and the twin multipole moment (TMM) method~\cite{Liu2025A}. We first simulate the Brownian motion of a passive helical flagellum. Subsequently, the analytical solutions for bacterial velocities are derived using the chiral TB model. Finally, the bacterial trajectories obtained by these simulation methods are quantitatively assessed using metrics such as the eigenvalues of the radius of gyration tensor\cite{Theodorou1985,Blavatska2010,Arkin2013}, the directionality ratio~\cite{Gorelik2014}, and the mean directional cosine of bacterial forward direction. Through this integrated approach, our aim is to investigate the critical role of flagellar morphology in determining the stability of bacterial Brownian motion and to robustly validate the effectiveness of the chiral TB model in simulating such motion.

\section{Model and Methods}
\label{sec2}
The rotation of flagella in a low Reynolds number fluid generates thrust, allowing the microorganism to swim. Since viscous forces are much greater than inertial forces of bacteria, the fluid dynamics is described by the incompressible Stokes equations:
\begin{equation}
\begin{split}
&\mu\nabla^{2}\textbf{u}-\nabla p=-\textbf{f},\\
&\nabla\cdot\textbf{u}=0.
\label{eq:refname01}
\end{split}
\end{equation}
where $\mu$ is the dynamic viscosity, $\textbf{u}$ is the fluid velocity, $p$ is the pressure, and $\textbf{f}$ is the external force per unit volume.

\begin{figure}
\centering
\includegraphics[width=0.7\textwidth]{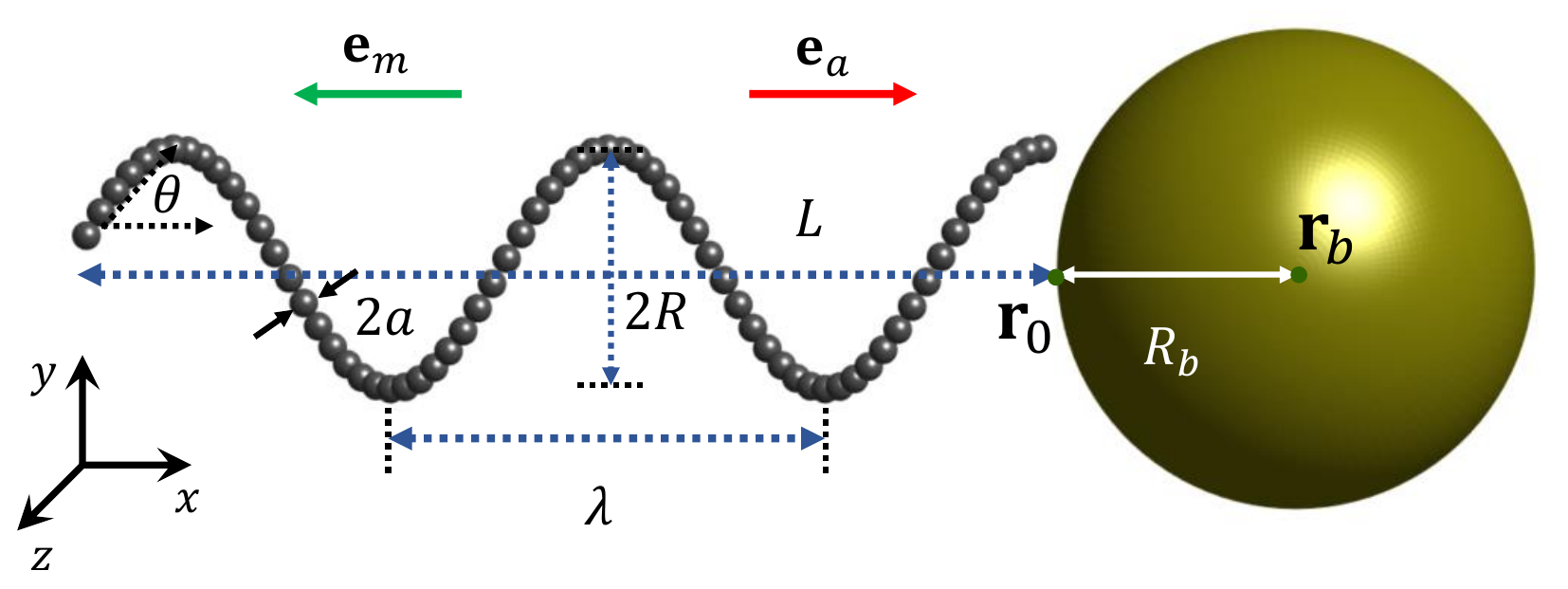}
\caption{Schematic diagram of a bacterium model. The flagellar axis is along the $x$-axis. The flagellum is characterized by a helix radius $R$, a filament radius $a$, a pitch $\lambda$, a pitch angle $\theta$, an axial length $L$, and a contour length $\Lambda=L/\cos\theta$, where $\tan\theta=2\pi R/\lambda$. The radius of the spherical cell body is denoted as $R_{b}$. The axial direction of the flagellum and the rotation direction of the motor are represented by $\mathbf{e}_{a}$ and $\mathbf{e}_{m}$, respectively.}
\label{fig:fig1}
\end{figure}

The bacterium is modeled as a sphere connected to a long, rigid, left-handed helical flagellum, as illustrated in Fig.~\ref{fig:fig1}. The radius of the cell body is denoted as $R_{b}$, with its center located at $\mathbf{r}_{b}$. The connection point between the flagellar axis and the cell body is represented by $\mathbf{r}_{0}$. The morphological parameters of the flagellum include a helix radius $R$, a pitch $\lambda$, a filament radius $a$, an axial length $L$, a pitch angle $\theta$ and the contour length $\Lambda=L/\cos\theta$, where $\tan\theta=2\pi R/\lambda$. In the reference frame of the flagellum, the centerline of the left-handed helical flagellum is expressed as:
\begin{equation}
\begin{split}
\mathbf{r}=\left[l\cos\theta,R\sin\varphi,R\cos\varphi\right].
\label{eq:refname02}
\end{split}
\end{equation}
where $l\in[0,\Lambda]$ is the contour length along the flagellum and $\varphi=\frac{2\pi}{\lambda}l\cos\theta$ represents the flagellar phase.

The linearity of the Stokes equations implies that the relationship between the forces and velocities of bacteria is linear. Specifically, when bodies exert forces $\mathbf{F}$ and torques $\mathbf{T}$ on the fluid, their translational velocities $\mathbf{U}$ and rotational velocities $\mathbf{W}$ satisfy the following equation:
\begin{equation}
\left(\begin{matrix}\mathbf{F}\\ \mathbf{T}\end{matrix}\right)=\mathcal{R}\left(\begin{matrix}\mathbf{U}-\mathbf{U}^{\infty}\\ \mathbf{W}-\mathbf{W}^{\infty}\end{matrix}\right).
\label{eq:refname03}
\end{equation}
where $\mathbf{U}^{\infty}$ and $\mathbf{W}^{\infty}$ are the ambient flow fields, while $\mathbf{U}-\mathbf{U}^{\infty}$ and $\mathbf{W}-\mathbf{W}^{\infty}$ represent the relative velocities of the bacterial system. The resistance matrix $\mathcal{R}$ used in the motion equations of the bacteria is computed individually using RFT~\cite{Gray1955,Chwang1975,Johnson1979}, the chiral TB model~\cite{Di2011}, and TMM~\cite{Liu2025A}. The chiral TB model is derived via RFT integration performed along the flagellar centerline. When the resistance matrix of the flagellum is computed using RFT, the hydrodynamic interactions between the cell body and the flagellum are neglected. Using TMM, hydrodynamic interactions between the flagellum and the cell body are effectively incorporated into the calculation of the bacterial resistance matrix. A comparison of the simulation results obtained from these three methods is conducted to validate the effectiveness of the chiral TB model in simulating the Brownian motion of bacteria.

\subsection{Resistance Force Theory}
\label{sec2.1}
The flagellar resistance matrix is computed using RFT, and is expressed as follows:
\begin{equation}
\begin{split}
\mathcal{R}_{t}=k_{\parallel}\hat{\mathbf{t}}\otimes\hat{\mathbf{t}}+k_{\perp}\left(\mathbb{I}-\hat{\mathbf{t}}\otimes\hat{\mathbf{t}}\right).
\label{eq:refname04}
\end{split}
\end{equation}
There are $N-1$ segments along the flagellum. The detailed expression for this $6(N-1)\times6(N-1)$ flagellar resistance matrix and the $6\times6$ resistance matrix $\mathcal{R}_{b}$ of the cell body can be found in Appendix A.

\subsection{Chiral Two-Body Model}
\label{sec2.2}
In the work of Leonardo et al.~\cite{Di2011}, RFT was used to calculate the resistance matrix of the flagellum. They integrated the influence of each flagellar segment into a single chiral body located at the center of the flagellum. This model simplifies the hydrodynamics of the flagellum by representing it as a force and torque exerted by this chiral body on the surrounding fluid, resulting in a $6\times6$ resistance matrix that incorporates the chirality and morphological characteristics. This approach significantly reduces computational costs. The helical flagellum is modeled as a chiral body and its resistance matrix is expressed as follows:
\begin{equation}
\begin{split}
&\mathbf{A}=X_{\parallel}^{A}\mathbf{e}_{a}\otimes\mathbf{e}_{a}+X_{\perp}^{A}\left(\mathbb{I}-\mathbf{e}_{a}\otimes\mathbf{e}_{a}\right),\\
&\mathbf{B}=X_{\parallel}^{B}\mathbf{e}_{a}\otimes\mathbf{e}_{a}+X_{\perp}^{B}\left(\mathbb{I}-\mathbf{e}_{a}\otimes\mathbf{e}_{a}\right),\\
&\mathbf{C}=X_{\parallel}^{C}\mathbf{e}_{a}\otimes\mathbf{e}_{a}+X_{\perp}^{C}\left(\mathbb{I}-\mathbf{e}_{a}\otimes\mathbf{e}_{a}\right).
\label{eq:refname05}
\end{split}
\end{equation}
where $\mathbf{e}_{a}$ is the axial direction of the flagellum, which is along the $x$-axis in this study. Detailed expressions for the components $X_{\parallel}^{A}$, $X_{\perp}^{A}$, $X_{\parallel}^{B}$, $X_{\perp}^{B}$, $X_{\parallel}^{C}$, and $X_{\perp}^{C}$ of the resistance matrix are provided in Appendix A. The resistance matrix of the bacterial system is:
\begin{equation}
\mathcal{R}=\left(\begin{matrix}\mathcal{R}_{b} & \mathbf{0}\\ \mathbf{0} & \mathcal{R}_{t}\end{matrix}\right).
\label{eq:refname06}
\end{equation}
where $\mathbf{0}$ is a $6\times6$ zero matrix. $\mathcal{R}$ is a $12\times12$ resistance matrix, and this method can significantly reduce computational costs, making it suitable for computationally intensive numerical simulations.

\subsection{Twin Multipole Moment}
\label{sec2.3}
The process of constructing the resistance matrix of the system using TMM is as follows~\cite{Durlofsky1987,Brady1988,Liu2025A}:
\begin{equation}
\begin{split}
\mathcal{R}=(\mathcal{M}^{\infty})^{-1}+\mathcal{R}_{2B,lub}.
\label{eq:refname07}
\end{split}
\end{equation}
where the subscript "lub" represents "lubrication". The grand resistance matrix incorporates both far-field hydrodynamic interaction, achieved by inverting $\mathcal{M}^{\infty}$, and pairwise lubrication interactions, represented by $\mathcal{R}_{2B,lub}$. The detailed procedure for calculating the resistance matrix using TMM has been provided in a previous work~\cite{Liu2025A}. The bacterial system consists of $N-1$ spheres in the flagellum and a spherical cell body.

\begin{table}
\caption{\label{tab:table1}Parameters of the numerical model and their corresponding values.}
\begin{tabular}{@{}p{3cm}p{6.5cm}l@{}}
\textbf{Notation} & \textbf{Description} & \textbf{Value}\\
\hline
$\mu$ & Dynamic viscosity & $1.0$~\si{\mu g/(\mu m\cdot s)}\\
$R_{b}$ & Radius of cell body & $1.0$~\si{\mu m}\\
$f$ & Motor rotation rate & $100$~\si{Hz}\\
$\mathbf{e}_{a}$ & Flagellar axis direction & $x$-axis\\
$\mathbf{e}_{m}$ & Rotation direction of motor & $\mathbf{e}_{m}=\frac{\mathbf{r}_{0}-\mathbf{r}_{b}}{|\mathbf{r}_{0}-\mathbf{r}_{b}|}$\\
$\mathbf{W}_{0}$ & Angular velocity of motor & $2\pi f\mathbf{e}_{m}$\\
$\mathbf{W}_{b}$ & Angular velocity of cell body & Variable\\
$\mathbf{W}_{t}$ & Angular velocity of flagellum & $\mathbf{W}_{t}=\mathbf{W}_{b}+\mathbf{W}_{0}$\\
$\mathbf{e}_{bact}$ & Forward direction of bacteria & $-\mathbf{W}_{t}/|\mathbf{W}_{t}|$\\
\hline
\end{tabular}
\end{table}

The motion of the cell body is described by a translational velocity $\textbf{U}_{b}$ and a rotational velocity $\mathbf{W}_{b}$. An arbitrary point $\mathbf{r}$ on the rigid flagellum experiences translational and rotational velocities as follows:
\begin{equation}
\begin{split}
&\mathbf{W}_{t}(\mathbf{r})=\mathbf{W}_{b}+\mathbf{W}_{0}\\
&\mathbf{U}_{t}(\mathbf{r})=\mathbf{U}_{b}+\mathbf{W}_{t}(\mathbf{r})\times\left(\mathbf{r}-\mathbf{r}_{b}\right).
\label{eq:refname08}
\end{split}
\end{equation}
where $\mathbf{W}_{0}=2\pi f\mathbf{e}_{m}$ is the angular velocity of the motor, and $\mathbf{r}$ and $\mathbf{r}_{b}$ are the position vectors of the flagellum and the center of the cell body, respectively. Assuming a fixed motor rotation rate of $f=100$~\si{Hz}, the unknowns $\mathbf{U}_{b}$ and $\mathbf{W}_{b}$ are determined using Eq.~\ref{eq:refname03}. The forward direction of the bacteria is defined as $\mathbf{e}_{bact}=-\mathbf{W}_{t}/|\mathbf{W}_{t}|$~\cite{Liu2025B}.

Neglecting inertia, we apply the force and torque balance conditions to the freely swimming bacterial system:
\begin{equation}
\begin{split}
&\mathbf{F}_{b}+\sum_{i=1}^{N-1}\mathbf{F}_{t}^{i}=0,\\
&\mathbf{T}_{b}+\sum_{i=1}^{N-1}\mathbf{T}_{t}^{i}+\sum_{i=1}^{N-1}\left(\mathbf{r}^{i}-\mathbf{r}_{b}\right)\times\mathbf{F}_{t}^{i}=0.
\label{eq:refname09}
\end{split}
\end{equation}
where $\textbf{F}_{b}$ and $\textbf{T}_{b}$ are the force and torque exerted on the fluid by the cell body, and $\textbf{F}_{t}^{i}$ and $\textbf{T}_{t}^{i}$ are the forces and torques exerted on the fluid by the $i$-th flagellar element, respectively. The parameters of the bacteria model are summarized in Table~\ref{tab:table1}. 

The resistance and mobility matrix of bacteria are denoted as $\mathcal{R}$ and $\mathcal{M}$, respectively. Since the off-diagonal elements of these matrices are significantly smaller than the diagonal elements, neglecting the off-diagonal elements simplifies the calculation of the flagellar stochastic forces and velocities~\cite{Li2008,Bafaluy1993,Schroeder2004,Martin2019}. Therefore, the stochastic velocities of the bacterial system can be approximately expressed as~\cite{Li2008}
\begin{equation}
\begin{split}
&\mathbf{U}^{B}=\sqrt{\frac{2k_{B}T}{dt}}\sqrt{\mathcal{M}_{ii}}\xi_{1}.
\label{eq:refname10}
\end{split}
\end{equation}
where $\xi_{1}$ is an independent Gaussian random variable with zero mean and unit variance. The constant $k_{B}$ is the Boltzmann constant, $T$ is the absolute temperature, and the Brownian time scale is $dt=10^{-6}$~\si{s}~\cite{Li2010}. The subscript "$ii$" indicates that only the diagonal elements of these matrices are considered.

\section{Results}
\label{sec3}
\subsection{Brownian Motion of a Helical Flagellum}
\label{sec3.1}
To assess the effectiveness of the chiral body model in simulating the Brownian motion of a helical flagellum, we compare the standard deviations of the translational and rotational velocities of the flagellar center predicted by the model with those obtained from RFT simulations. The morphological parameters of the flagellum are as follows: filament radius $a=0.04$~\si{\mu m}, pitch angle $\theta=\pi/5$, helix radius $R=0.2$~\si{\mu m}, and contour length $\Lambda=7.0$~\si{\mu m}. 

\begin{figure}
\centering
\includegraphics[width=0.7\textwidth]{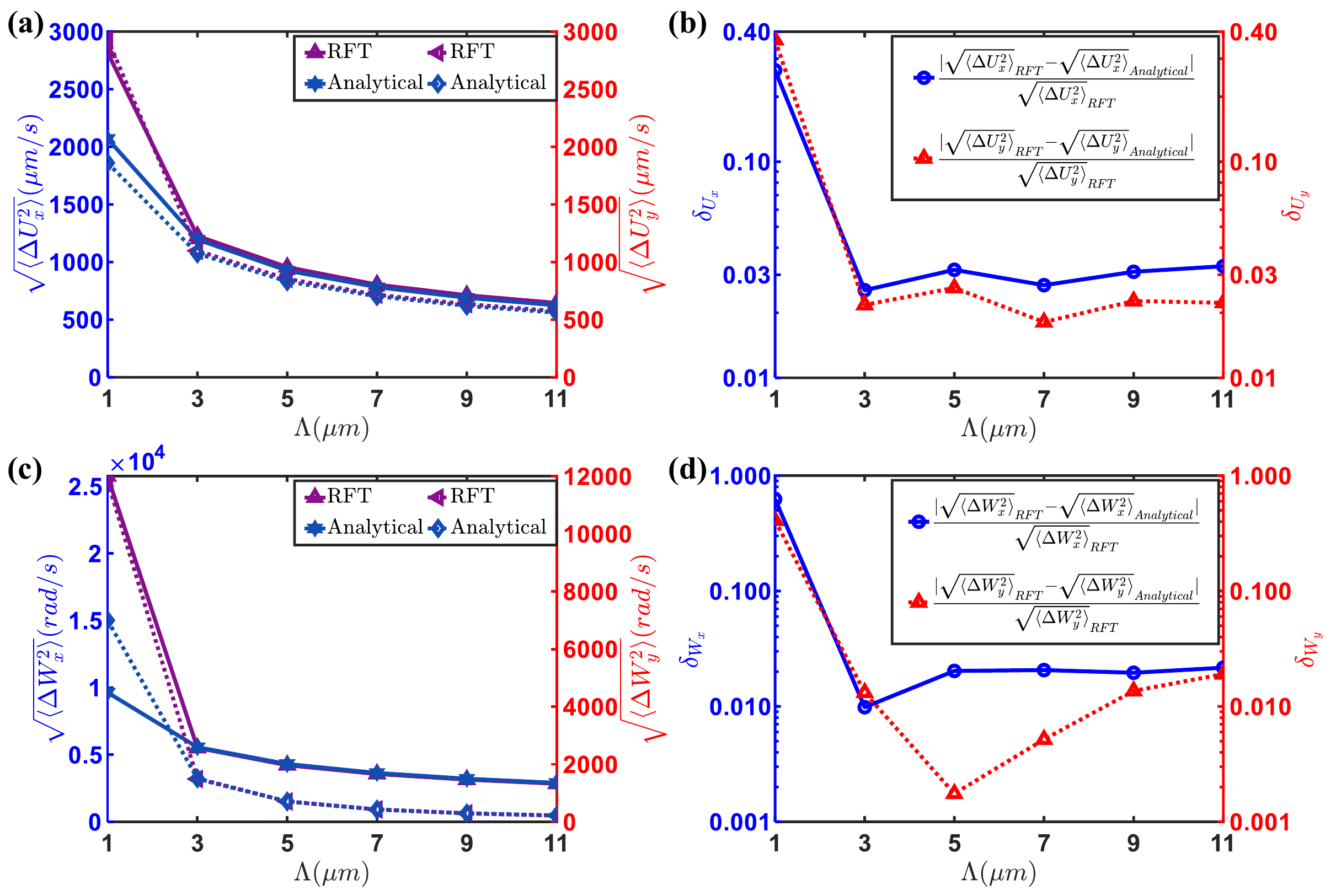}
\caption{(a) and (c) Standard deviations of the translational and rotational velocities of the flagellar center as a function of contour length $\Lambda$, respectively. Solid lines represent quantities along the $x$-axis, while the dashed lines represent quantities along the $y$-axis. Correspondingly, (b) and (d) Relative differences between the analytical solutions and RFT simulations for these translational and rotational velocity standard deviations.}
\label{fig:fig2}
\end{figure}

\begin{figure}
\centering
\includegraphics[width=0.7\textwidth]{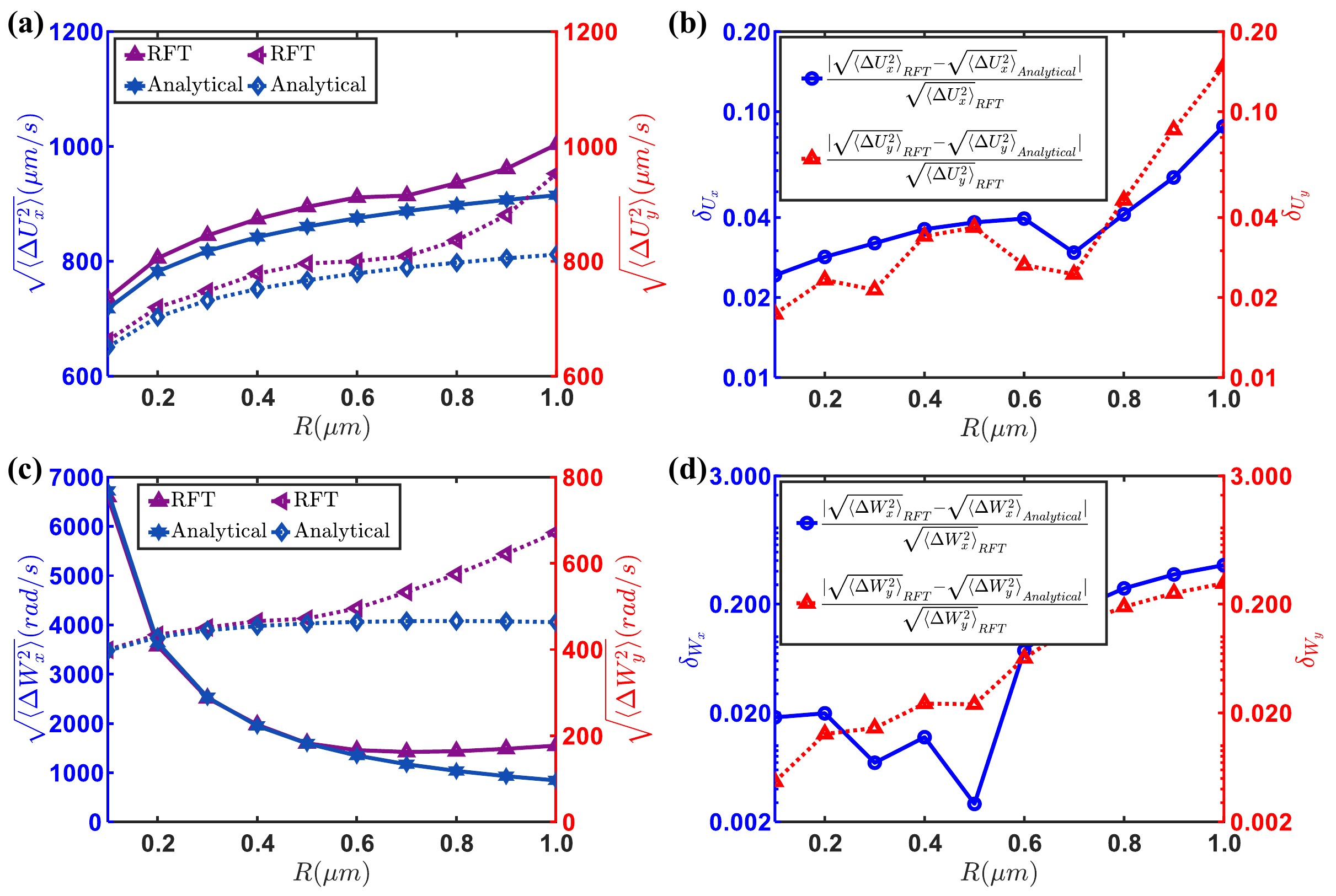}
\caption{Standard deviations of the flagellar center's (a) translational and (c) rotational velocities as a function of helix radius $R$. The solid and dashed lines represent the $x$-axis and $y$-axis quantities, respectively. (b) and (d) Relative differences between the analytical solutions and RFT simulations for these translational and rotational velocity standard deviations, respectively.}
\label{fig:fig3}
\end{figure}

In this study, the flagellar axis is along the $x$-axis. Phase averaging is performed on the helical flagellum, resulting in axial symmetry. Consequently, our analysis focuses only on translational and rotational velocities along the $x$-axis and $y$-axis. The contour length $\Lambda$ and helix radius $R$ are two critical morphological parameters of the flagellum. As illustrated in Figs.~\ref{fig:fig2} and~\ref{fig:fig3}, the standard deviations of the translational and rotational velocities of the flagellar center along the $x$-axis and $y$-axis are computed using RFT. Furthermore, analytical expressions for these standard deviations, given by $\sqrt{2k_{B}T/dt}\sqrt{\mathcal{M}_{ii}}$, are derived from the chiral body model, specifically Eqs.~\ref{eq:refname05} and \ref{eq:refname10}, and these solutions are then compared with RFT simulations to analyze the relative differences. All results are averaged over $10^{5}$ simulation runs.

Figs.~\ref{fig:fig2}(a) and (c) demonstrate that, with increasing flagellar contour length, the standard deviations of both translational and rotational velocities of the flagellar center along the $x$-axis and $y$-axis decrease, suggesting that the stability of the flagellar Brownian motion is improved. Figs.~\ref{fig:fig2}(b) and (d) illustrate the relative differences between the standard deviations obtained from the chiral body model and those from RFT. In Fig.~\ref{fig:fig2}(b), the blue solid line represents the relative difference of the standard deviations of the translational velocities along the $x$-axis, which remains below $4\%$ for $\Lambda\ge3.0$~\si{\mu m}. The red dashed line indicates the relative difference of the standard deviations of the translational velocity along the $y$-axis, staying below $3\%$ when $\Lambda\ge3.0$~\si{\mu m}. Similarly, in Fig.~\ref{fig:fig2}(d), the blue solid line shows the relative difference of the standard deviations of the rotational velocities along the $x$-axis, which remains below $3\%$ when $\Lambda\ge3.0$~\si{\mu m}. The red dashed line illustrates the relative difference of the standard deviations of the rotational velocities along the $y$ axis, which is less than $2\%$ for $\Lambda\ge3.0$~\si{\mu m}. In conclusion, for $\Lambda\ge3.0$~\si{\mu m}, the chiral body model effectively simulates the Brownian motion of the flagellum.

Fig.~\ref{fig:fig3} shows the standard deviations of the translational and rotational velocities of the flagellar center as functions of the helix radius $R$. Figs.~\ref{fig:fig3}(a) and (c) reveal that the results obtained from the chiral body model closely agree with those from RFT for $R\le0.5$~\si{\mu m}. Figs.~\ref{fig:fig3}(b) and (d) show the relative differences between the standard deviations obtained from the chiral body model and the RFT simulations for translational and rotational velocities, respectively. Specifically, Fig.~\ref{fig:fig3}(b) illustrates these relative differences for the translational velocities of the flagellar center. For $R\le0.7$~\si{\mu m}, these relative differences are within $4\%$, indicating a good consistency between the two models in simulating the flagellar Brownian motion. Similarly, for $R\le0.5$~\si{\mu m}, the relative differences in the standard deviations of the rotational velocities along both axes remain below $3\%$. In summary, within the range of $R\le0.5$~\si{\mu m}, the chiral body model effectively simulates the flagellar Brownian motion.

\begin{figure}
\centering
\includegraphics[width=0.70\textwidth]{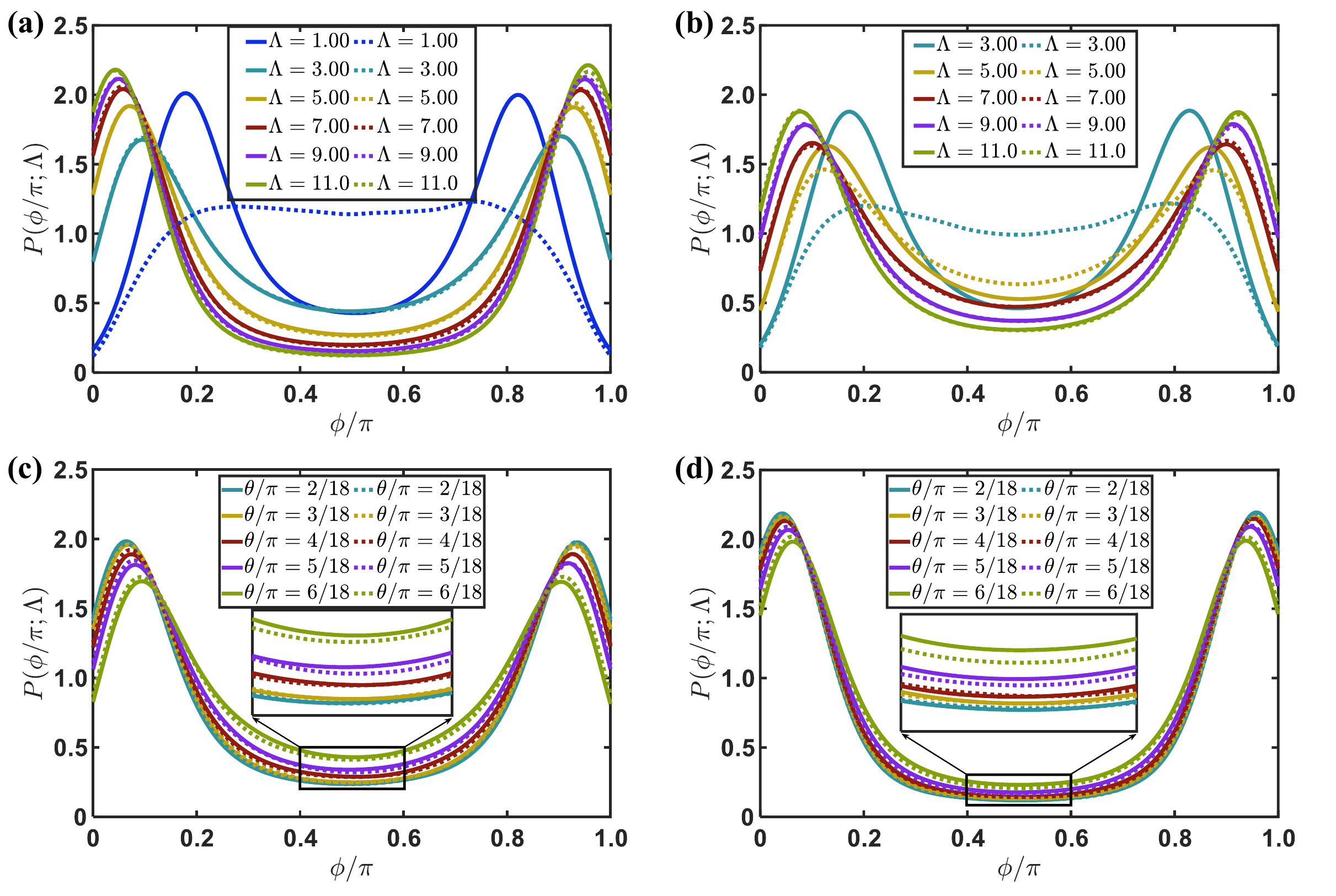}
\caption{Probability distribution of the angle $\phi$ for different contour lengths at helix radii (a) $R=0.20$~\si{\mu m} and (b) $R=0.50$~\si{\mu m}, respectively. Probability distribution of the angle $\phi$ for different pitch angles for contour lengths (c) $\Lambda=5.00$~\si{\mu m} and (d) $\Lambda=10.0$~\si{\mu m}, respectively. The solid lines represent the results obtained from RFT simulations, while dashed lines correspond to those from the chiral body model.}
\label{fig:fig4}
\end{figure}

In a viscous fluid, the Brownian motion of a flagellum induces randomness in its rotational velocity, affecting both its transverse and longitudinal components. This stochasticity makes it challenging to definitively determine the direction of rotation. To analyze the flagellar rotation direction more rigorously, we introduce the probability distribution function $g(\phi)$ for angle $\phi$. Here, $\phi$ is defined as the angle between the flagellar rotational velocity vector $\mathbf{W}$ and the flagellar axis. This angle can be related to the components of $\mathbf{W}=\{W_{x},W_{y},W_{z}\}$ by $\sin\phi=\frac{\sqrt{W_{y}^{2}+W_{z}^{2}}}{\sqrt{W_{x}^{2}+W_{y}^{2}+W_{z}^{2}}}$. The probability distribution $g(\phi)$ satisfies the following normalization condition:
\begin{equation}
\begin{split}
\int_{0}^{\pi}g(\phi)\sin\phi d\phi=1.
\label{eq:refname11}
\end{split}
\end{equation}
In numerical simulations, we can only obtain the probability distribution function in the form of $P(\phi)=g(\phi)\sin\phi$. Considering the isotropy of a passive sphere that undergoes Brownian motion, the probability distribution function $g(\phi)$ can be assumed to be constant, specifically $g(\phi)=\frac{1}{2}$. Therefore, we obtain the following equation:
\begin{equation}
\begin{split}
P(\bar{\phi})=\frac{\pi}{2}\sin(\bar{\phi}).
\label{eq:refname12}
\end{split}
\end{equation}
where $\bar{\phi}=\phi/\pi$. Both the chiral body model and the RFT are used to simulate the probability distribution of the angle $\phi$. The angular probability density function $g(\phi)$ can be expressed as $g(\phi)=P(\phi)/\sin\phi$. However, singularities may occur as $\sin\phi$ approaches zero when $\phi$ approaches $0$ or $\pi$. Therefore, we use the function $P(\phi)$ to represent the probability distribution of the angle $\phi$. As shown in Fig.~\ref{fig:fig4}(a), for a helix radius of $R=0.20$~\si{\mu m} and a contour length of $\Lambda=3.0$~\si{\mu m}, the probability distribution obtained from the chiral body model is in close agreement with that from RFT simulations. This agreement improves as the contour length of the flagellum increases. Moreover, as depicted in Fig.~\ref{fig:fig4}(b), strong agreement is observed for $\Lambda\ge 5.0$~\si{\mu m} when the helix radius is $R=0.50$~\si{\mu m}.

When the flagellar contour length is $\Lambda=5.0$~\si{\mu m} and the helix radius is $R=0.20$~\si{\mu m}, Fig.~\ref{fig:fig4}(c) shows that the results of the chiral body model and RFT are in good agreement for pitch angles $\theta\le\frac{5}{18}\pi$. Similarly, Fig.~\ref{fig:fig4}(d) demonstrates that, for a contour length of $\Lambda=10.0$~\si{\mu m} and a helix radius of $R=0.20$~\si{\mu m}, the probability distribution function of $\phi$ remains highly consistent for $\theta\le\frac{5}{18}\pi$. Considering the four panels presented in Fig.~\ref{fig:fig4}, we find that when the helix radius $R\le0.5$~\si{\mu m}, the contour length $\Lambda\ge 5.0$~\si{\mu m}, and the pitch angles are in the interval $[\pi/6,2\pi/9]$, the distribution function $P(\phi)$ obtained from the chiral body model and RFT exhibit high consistency. From Fig.~\ref{fig:fig4}, it can be observed that the flagellar rotation velocity $\mathbf{W}$ has a maximum probability of being aligned with the flagellar axis.

\subsection{Non-Brownian Motion of a Bacterium}
\label{sec3.2}
Since RFT neglects hydrodynamic interactions between the cell body and the flagellum, the derived chiral TB model also does not include these interactions. However, previous studies have shown that these interactions can significantly influence the magnitude of thrust and torque~\cite{Darnton2007,Liu2025B}. To evaluate the influence of these hydrodynamic interactions, we simulate the bacterial thrust and torque in non-Brownian motion using the chiral TB model and the TMM. This section presents analytical solutions for bacterial translational and rotational velocities under non-Brownian conditions, derived from the chiral TB model. It also details the corresponding thrust and torque exerted by the motor on the cell body. The motor rotation velocity is defined as $\mathbf{W}_{0}=W_{0}\mathbf{e}_{m}=2\pi f\mathbf{e}_{m}$, where $f=100$~\si{Hz} is the motor rotation rate, and $\mathbf{e}_{m}$ denotes the rotational direction, assumed to be aligned with the flagellar axis for simplicity. The morphological parameters of the flagellum are set as follows: filament radius $a=0.01$~\si{\mu m}, pitch angle $\theta=\pi/5$, helix radius $R=0.25$~\si{\mu m}, and contour length $\Lambda=8.0$~\si{\mu m}. These values are similar to those of \emph{Escherichia coli}~\cite{Darnton2007}. Using the chiral TB model, bacterial motion is simulated assuming no thermal noise. Based on the equation of motion (Eq.~\ref{eq:refname03}) and the force and torque balance conditions (Eq.~\ref{eq:refname09}), the governing equations for bacterial motion along the $x$-axis and $y$-axis are expressed as:
\begin{equation}
\begin{split}
&\left(\begin{matrix} X_{\parallel}^{A}+6\pi\mu R_{b} & X_{\parallel}^{B}\\
X_{\parallel}^{B} & X_{\parallel}^{C}+8\pi\mu R_{b}^{3}\end{matrix}\right)
\left(\begin{matrix} U_{x} \\ W_{x}\end{matrix}\right)=
\left(\begin{matrix} -X_{\parallel}^{B}W_{0} \\ -X_{\parallel}^{C}W_{0} \end{matrix}\right),\\
&\left(\begin{matrix} X_{\perp}^{A}+6\pi\mu R_{b} & X_{\perp}^{B}\\
X_{\perp}^{B} & X_{\perp}^{C}+8\pi\mu R_{b}^{3}\end{matrix}\right)
\left(\begin{matrix} U_{y} \\ W_{y}\end{matrix}\right)=
\left(\begin{matrix} 0 \\ 0 \end{matrix}\right).
\label{eq:refname13}
\end{split}
\end{equation}

\begin{figure}
\centering
\includegraphics[width=0.8\textwidth]{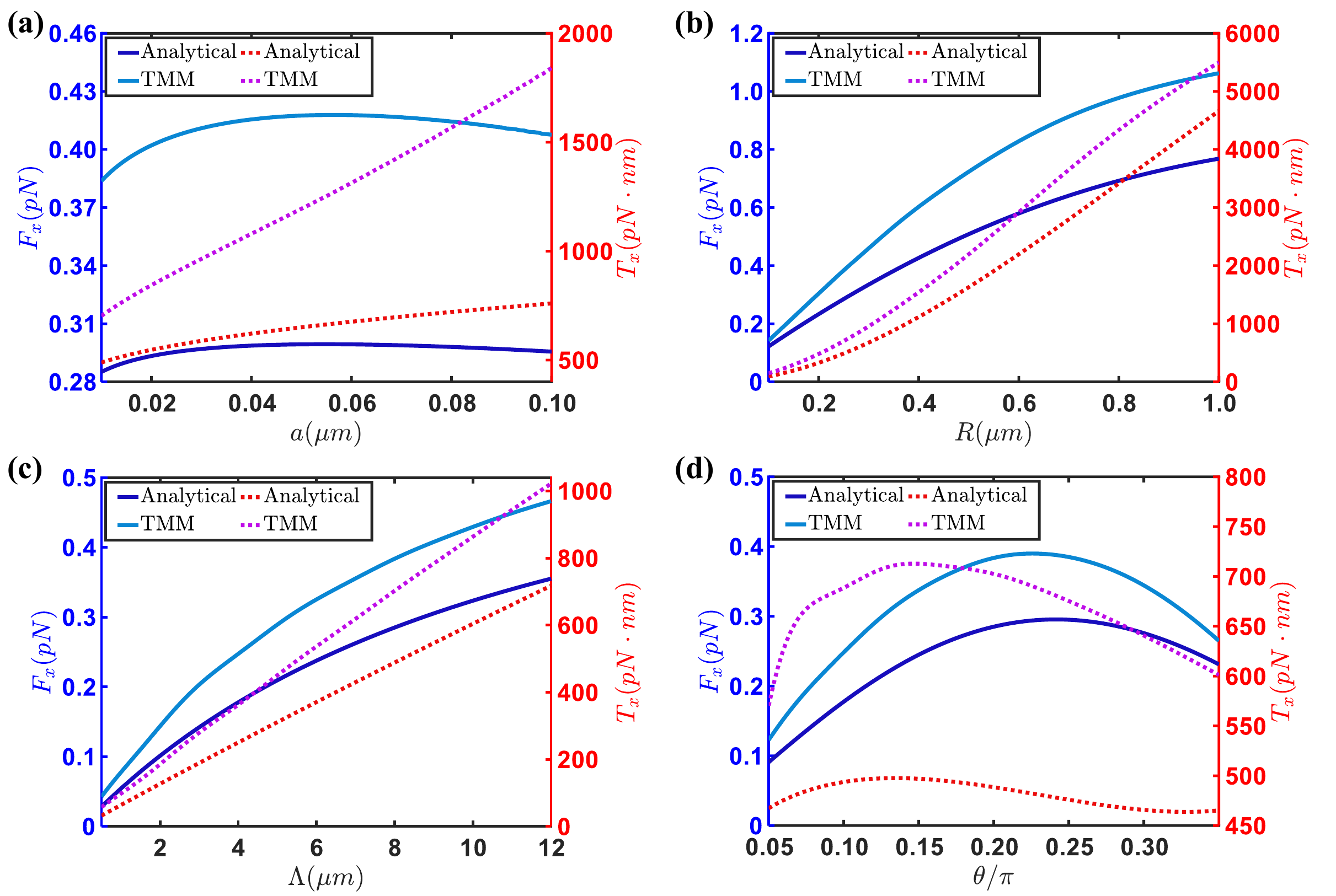}
\caption{Thrust and torque exerted by the motor on the cell body, obtained from the analytical solutions and the TMM simulations, depend on the following parameters: (a) filament radius $a$, (b) helix radius $R$, (c) contour length $\Lambda$, and (d) pitch angle $\theta$. The solid lines represent the thrust, while dashed lines represent the torque.}
\label{fig:fig5}
\end{figure}

Due to symmetry, the equations for the translational and rotational velocities of the bacterium along the $z$-axis are identical to those along the $y$-axis. Consequently, according to Eq.~\ref{eq:refname13}, the translational and rotational velocities of the cell body are expressed as follows:
\begin{equation}
\begin{split}
&U_{x}=\frac{8\pi\mu R_{b}^{3}X_{\parallel}^{B}}{(X_{\parallel}^{B})^{2}-(X_{\parallel}^{A}+6\pi\mu R_{b})(X_{\parallel}^{C}+8\pi\mu R_{b}^{3})}W_{0},\\
&W_{x}=\frac{X_{\parallel}^{A}X_{\parallel}^{C}-(X_{\parallel}^{B})^{2}+6\pi\mu R_{b}X_{\parallel}^{C}}{(X_{\parallel}^{B})^{2}-(X_{\parallel}^{A}+6\pi\mu R_{b})(X_{\parallel}^{C}+8\pi\mu R_{b}^{3})}W_{0},\\
&U_{y}=U_{z}=0,\\
&W_{y}=W_{z}=0.
\label{eq:refname14}
\end{split}
\end{equation}
where $W_{0}=-200\pi$. Eq.~\ref{eq:refname14} shows that the bacterium moves with translational and rotational velocities only along the flagellar axis, with zero velocities in all other directions. This implies that the trajectories obtained from the chiral TB model are straight lines. Eqs.~\ref{eq:refname14}, \ref{eq:refnameA05} and \ref{eq:refnameA10} reveal that these velocities are directly proportional to the motor rotation rate and independent of the dynamic viscosity. The thrust and torque exerted by the motor on the cell body are given:
\begin{equation}
\begin{split}
&F_{x}=6\pi\mu R_{b}U_{x},\\
&T_{x}=8\pi\mu R_{b}^{3}W_{x}.
\label{eq:refname15}
\end{split}
\end{equation}
Similarly, thrust and torque are proportional to the motor rotation rate and do not depend on dynamic viscosity. However, the relationships between thrust, torque, and the morphological parameters of the flagellum are more complex. To intuitively understand these relationships, we calculate the thrust and torque for various flagellar morphologies by varying the filament radius $a$, the helix radius $R$, the contour length $\Lambda$, and the pitch angle $\theta$, as illustrated in Fig.~\ref{fig:fig5}.

Fig.~\ref{fig:fig5} presents the thrust and torque obtained from analytical solutions in Eqs.~\ref{eq:refname14} and~\ref{eq:refname15}, as well as from TMM simulations. As shown in Figs.~\ref{fig:fig5}(a) and (d), there is no clear correlation between the filament radius and the pitch angle. However, within the ranges of $0.10\le R\le 0.30$~\si{\mu m} and $6.0\le\Lambda\le9.0$~\si{\mu m}, both the thrust and the torque are approximately proportional to the helix radius and the contour length of the flagellum, as illustrated in Figs.~\ref{fig:fig5}(b) and (c). These dimensions are similar to the morphological characteristics observed in typical bacteria used in experiments~\cite{Darnton2007,Koyasu1984,Fujii2008,Spagnolie2011}.

The thrust and torque exerted by the motor on the cell body, derived from the chiral TB model, are lower than those obtained by the TMM simulations. This discrepancy arises from RFT neglecting the hydrodynamic interactions between the different flagellar segments, as well as between the flagellum and the cell body. Previous studies have shown that these hydrodynamic interactions significantly influence bacterial motility~\cite{Liu2025B}. Experimental measurements indicate that the thrust exerted by the motor on the cell body of \emph{Escherichia coli} is about $F_{x}=0.32\pm0.08$~\si{pN}, and the torque is approximately $T_{x}=840\pm360$~\si{pN\cdot nm}~\cite{Darnton2007}. Using the TMM with the morphological parameters provided in this section, the calculated thrust and torque for \emph{Escherichia coli} are approximately $F_{x}\approx0.38$~\si{pN} and $T_{x}\approx700$~\si{pN\cdot nm}, respectively. Similarly, the thrust and torque calculated using RFT are about $F_{x}\approx0.28$~\si{pN} and $T_{x}\approx490$~\si{pN\cdot nm}, respectively. All of these values obtained from the RFT and TMM are within the range observed from the experiments. Although RFT neglects the hydrodynamic interactions between the cell body and the flagellum, it is widely used because of its low computational cost and the ability to incorporate the detailed morphology of the flagellum. Meanwhile, the chiral TB model derived from RFT can also achieve satisfactory performance in such applications.

\subsection{Brownian Motion of a Bacterium}
\label{sec3.3}
Bacteria swimming in fluid media are inevitably affected by thermal noise. Experiments have shown that colloids exhibit random walks, whereas bacteria demonstrate pronounced directional persistence, highlighting the critical role of flagella in maintaining the stability and directionality of bacterial motility. The chiral TB model simplifies a bacterium into a cell body and a chiral body that represent the flagellum, each described by a $6\times6$ resistance matrix. Without considering noise effects, the $6\times6$ resistance matrix of the flagellum can be derived by integrating the centerline and averaging the phases using RFT~\cite{Di2011}, which eliminates the non-axisymmetric components. Consequently, when transverse velocity is absent (indicating straight-line motion), the longitudinal translational velocity, rotational velocity, thrust, and torque exerted by the motor on the cell body, as calculated by the chiral TB model, are consistent with those obtained from RFT. However, the validity of the chiral TB model for simulating bacterial Brownian motion under thermal noise requires further validation. Numerical simulations are performed using the chiral TB model, RFT, and TMM to study the Brownian motion of bacteria. The parameters of the flagellum are as follows: filament radius $a=0.04$~\si{\mu m}, pitch angle $\theta=\pi/5$, helix radius $R=0.2$~\si{\mu m}, and contour length $\Lambda=6.0$~\si{\mu m}.

\begin{figure}
\centering
\includegraphics[width=1.0\textwidth]{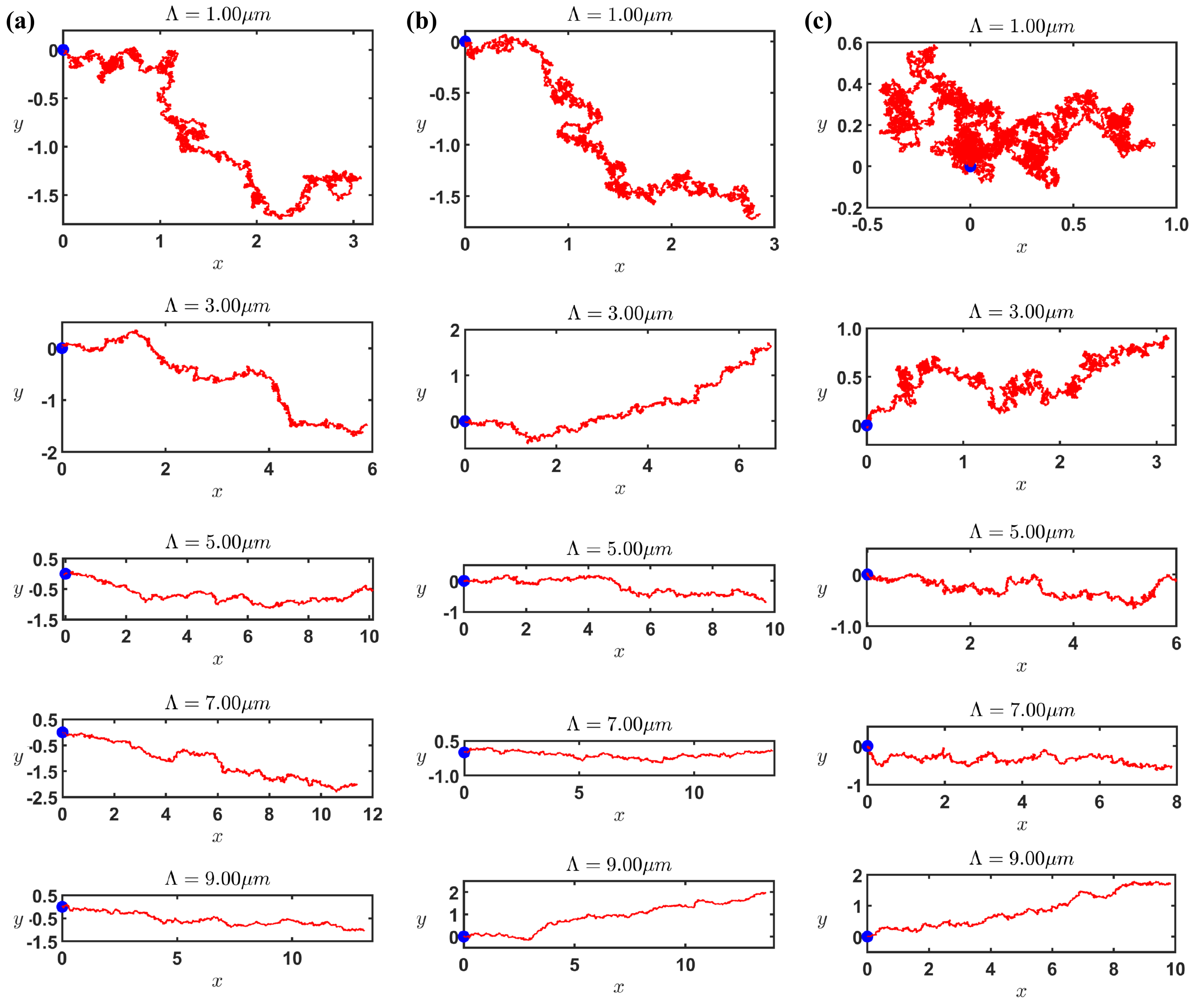}
\caption{Trajectories of bacterial Brownian motion over one second interval. Trajectories of bacteria with different flagellar contour lengths are simulated using: (a) the chiral TB model, (b) RFT, and (c) TMM. The blue dots represent the initial positions of the bacteria. All axes are in \si{\mu m}. The $x$-axis is along the flagellar axis, while the $y$-axis represents an arbitrary direction perpendicular to it.}
\label{fig:fig6}
\end{figure}

We simulate bacterial Brownian motion with different flagellar contour lengths using the chiral TB model, RFT, and TMM. The simulation algorithm is based on previous work~\cite{Liu2025B}. Figs.~\ref{fig:fig6}(a), (b) and (c) display representative bacterial trajectories obtained from these approaches, respectively. Bacteria with longer flagella tend to follow more linear paths, qualitatively consistent with experimental measurements~\cite{Darnton2007,Son2013,Bianchi2017,Figueroa2020,Grognot2021}. The trajectories obtained from the chiral TB model and the RFT exhibit similar shapes but differ slightly from those obtained from the TMM. This discrepancy arises mainly because the chiral TB model and the RFT do not consider hydrodynamic interactions between the cell body and the flagellum, while the TMM includes these interactions. The results indicate that increasing the contour length of the flagella enhances the directionality of bacterial motion. However, Fig.~\ref{fig:fig6} presents trajectories qualitatively and does not quantitatively evaluate the differences and stability of these trajectories.

To quantitatively assess the effect of flagellar morphology on the stability and directionality of bacterial Brownian motion, we introduce several quantitative metrics: the radius of gyration tensor of trajectories, the directionality ratio, and the mean directional cosine of the bacterial forward direction. The radius of gyration tensor is calculated as follows~\cite{Theodorou1985,Blavatska2010,Arkin2013}:
\begin{equation}
\begin{split}
S=\frac{1}{N_{t}}\sum_{i=1}^{N_{t}}\left(\mathbf{X}_{i}-\mathbf{X}_{cm}\right)\otimes\left(\mathbf{X}_{i}-\mathbf{X}_{cm}\right).
\label{eq:refname16}
\end{split}
\end{equation}
where $\mathbf{X}_{i}$ represents the position at the $i$-th time step in the bacterial trajectory, $\mathbf{X}_{cm}$ denotes the center of the trajectory and $N_{t}$ is the total number of time steps. The eigenvalues of the radius of gyration tensor $S$, denoted as $\lambda_{1}$, $\lambda_{2}$, and $\lambda_{3}$, serve as shape descriptors for bacterial trajectories, ordered in ascending sequence such that $\lambda_{1} \leq \lambda_{2} \leq \lambda_{3}$~\cite{Theodorou1985,Blavatska2010,Arkin2013}.

\begin{figure}
\centering
\includegraphics[width=0.8\textwidth]{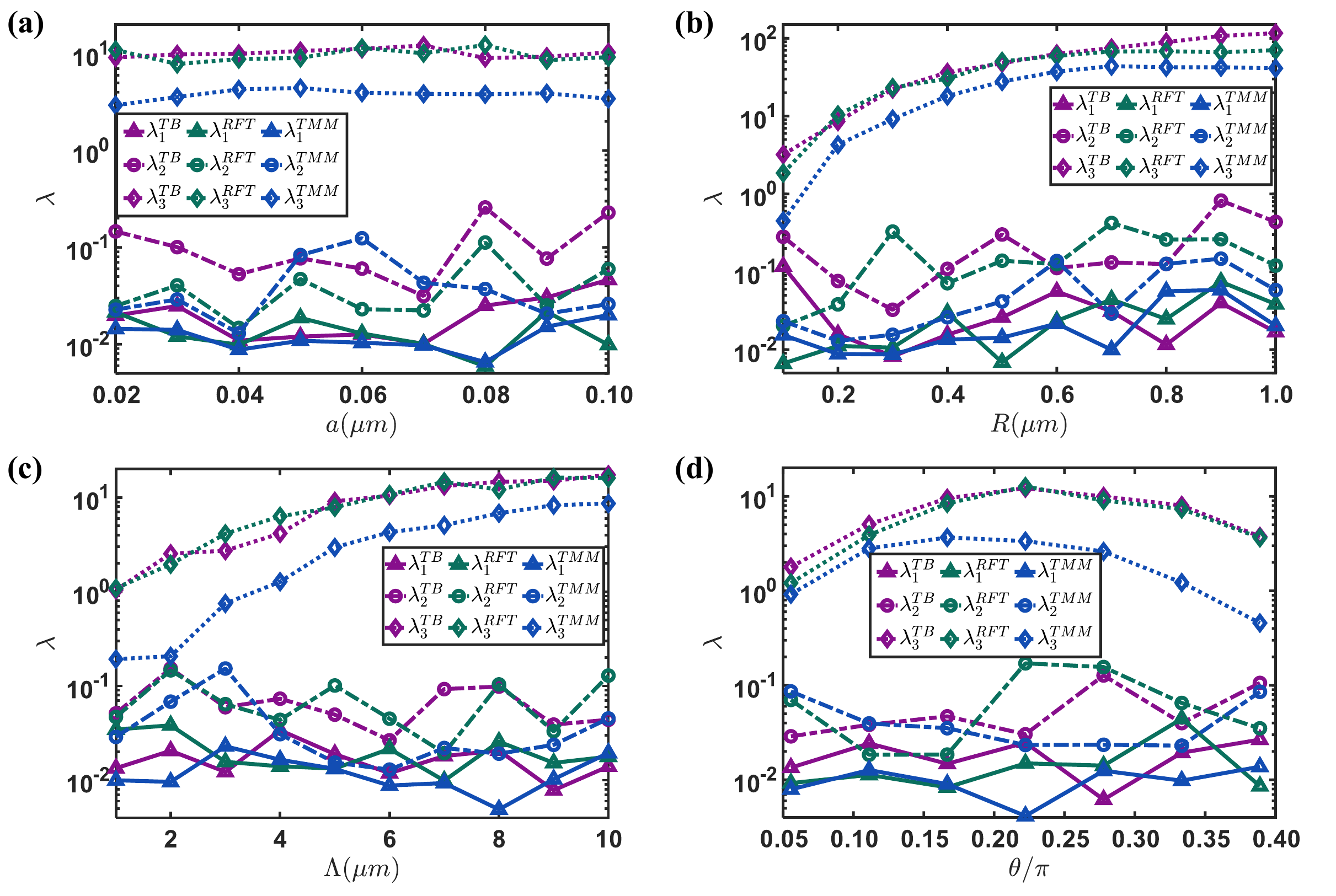}
\caption{Eigenvalues of the radius of gyration tensor for bacterial trajectories as functions of (a) filament radius $a$, (b) helix radius $R$, (c) contour length $\Lambda$, and (d) pitch angle $\theta$. These eigenvalues are computed using the chiral TB model, RFT, and TMM, respectively.}
\label{fig:fig7}
\end{figure}

The eigenvalues $\lambda_{1}$, $\lambda_{2}$, and $\lambda_{3}$ of the tensor $S$ are illustrated in Fig.~\ref{fig:fig7}. These eigenvalues of the radius of gyration tensor characterize the symmetry and elongation of bacterial trajectories. Larger eigenvalues indicate greater elongation of the trajectories. When $\lambda_{3}$ is significantly larger than $\lambda_{1}$ and $\lambda_{2}$, it suggests that the bacterial trajectory is highly elongated and exhibits strong directionality. In the case of predominantly linear motion, the ratios $\lambda_{3}/\lambda_{1}$ and $\lambda_{3}/\lambda_{2}$ approach infinity, while for symmetric spherical trajectories, these ratios are equal to $1$. However, these ratios are overly sensitive to the smallest eigenvalues, potentially leading to significant errors. Therefore, we use the eigenvalues themselves rather than their ratios for characterization.

As shown in Fig.~\ref{fig:fig7}(a), the filament radius $a$ has little effect on the shape of bacterial trajectories. The eigenvalues of the radius of gyration tensor, obtained from simulations using the chiral TB model, RFT, and TMM, are comparable in magnitude. This indicates that the chiral TB model effectively simulates bacterial Brownian motion with low sensitivity to variations in filament radius.

Fig.~\ref{fig:fig7}(b) shows that with an increase in helix radius, $\lambda_{3}$ increases significantly, while the other two eigenvalues show negligible change. This suggests that increasing the helix radius results in the elongation of the trajectory. The simulation results obtained from the chiral TB model, RFT, and TMM indicate that, except for the case $R=0.10$~\si{\mu m}, the eigenvalues of the radius of gyration tensor are of the same order of magnitude for all helix radii. Consequently, for $R\ge0.2$~\si{\mu m}, the chiral TB model provides reliable results in simulating the bacterial Brownian motion.

As illustrated in Fig.~\ref{fig:fig7}(c), as the contour length $\Lambda$ increases, the value of $\lambda_{3}$ also increases, while the other two eigenvalues remain relatively constant. The increased contour length markedly enhances the elongation of bacterial trajectories. Overall, the eigenvalues obtained from the chiral TB model are consistent in magnitude with those obtained from RFT and TMM for $\Lambda\ge 4.0$~\si{\mu m}. However, when $\Lambda<4.0$~\si{\mu m}, the results obtained from the chiral TB model agree well with those from RFT but diverge significantly from TMM.

The simulation results from the chiral TB model remain highly consistent with those from RFT as the pitch angle varies, as shown in Fig.~\ref{fig:fig7}(d). However, when the pitch angle $\theta$ exceeds $2\pi/9$, significant discrepancies arise between the values of $\lambda_{3}$ obtained from these two methods and those from TMM. This is because RFT only accounts for the local hydrodynamic interactions of the flagellum, which has been validated~\cite{Rodenborn2013,Liu2025A}. A comparison between the chiral TB model and the RFT results shows that both remain within the same order of magnitude, further confirming the effectiveness of the chiral TB model in simulating bacterial Brownian motion.

In the absence of thermal noise, bacterial trajectories take the form of cylindrical helices. Although the eigenvalues of the radius of gyration tensor effectively characterize the shape of the trajectory, they do not quantitatively capture the ratio of displacement to trajectory length. Since the trajectory length is typically greater than the displacement for helical trajectories, we introduce the directionality ratio as a metric to quantify the linearity of bacterial motility. It is defined as the straight-line distance between the starting and ending points of the trajectory divided by the total trajectory length~\cite{Gorelik2014}. This ratio approaches $1$ for linear bacterial trajectories, while it approaches $0$ for highly curved trajectories.

\begin{figure}
\centering
\includegraphics[width=0.8\textwidth]{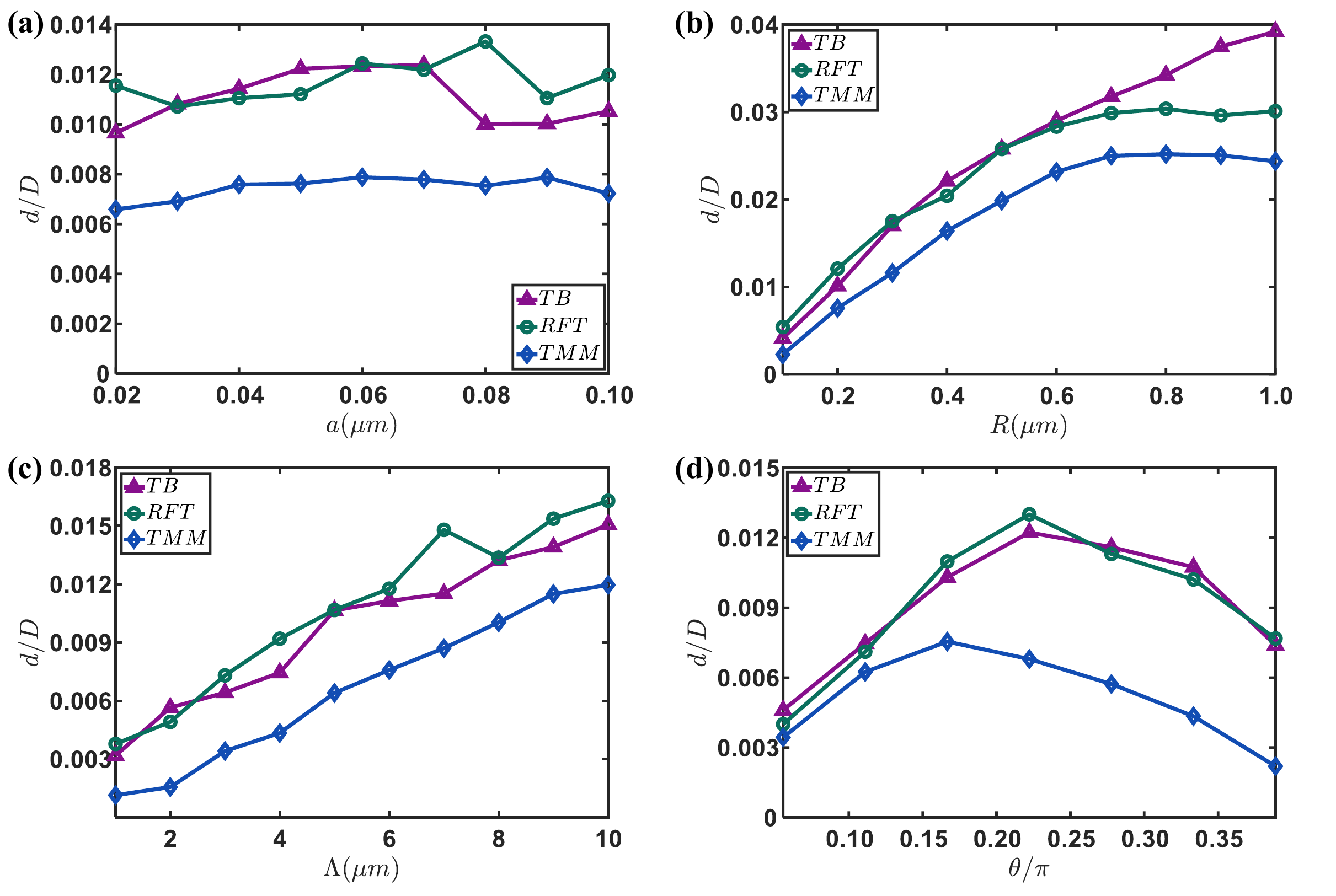}
\caption{Directionality ratio of bacterial trajectories as functions of (a) filament radius $a$, (b) helix radius $R$, (c) contour length $\Lambda$, and (d) pitch angle $\theta$, obtained from the chiral TB model, RFT, and TMM, respectively.}
\label{fig:fig8}
\end{figure}

The influence of the filament radius of the flagella on the directionality ratio is negligible, as shown in Fig.~\ref{fig:fig8}(a). Figs.~\ref{fig:fig8}(b) and (c) demonstrate that the directionality ratio increases with increasing helix radius and contour length of the flagella. Furthermore, Fig.~\ref{fig:fig8}(d)  reveals a specific pitch angle at which the directionality ratio reaches its maximum, corresponding to the maximum value of $\lambda_{3}$ shown in Fig.~\ref{fig:fig7}(d). When analyzing $\lambda_{3}$ from Fig.~\ref{fig:fig7} and the directionality ratio in Fig.~\ref{fig:fig8}, we found that the values predicted by the chiral TB model are similar to those obtained from RFT, and both are slightly higher than those obtained using TMM.

Rigid bacteria maintain a constant forward direction while swimming in fluid media in the absence of thermal noise~\cite{Liu2025B}. However, their trajectories typically exhibit cylindrical helical paths. Therefore, an appropriate metric is required to quantitatively characterize the directionality of bacterial motility. To quantify the deviation of the forward direction of the bacteria from its initial flagellar axis, we define the mean directional cosine as $\langle\mathbf{e}_{bact}(t)\cdot\mathbf{e}_{a}\rangle$, where $\mathbf{e}_{bact}(t)=-\mathbf{W}_{t}(t)/|\mathbf{W}_{t}(t)|$ represents the instantaneous forward direction of the bacteria and $\mathbf{e}_{a}$ denotes the initial flagellar axial direction~\cite{Liu2025B}.

\begin{figure}
\centering
\includegraphics[width=0.8\textwidth]{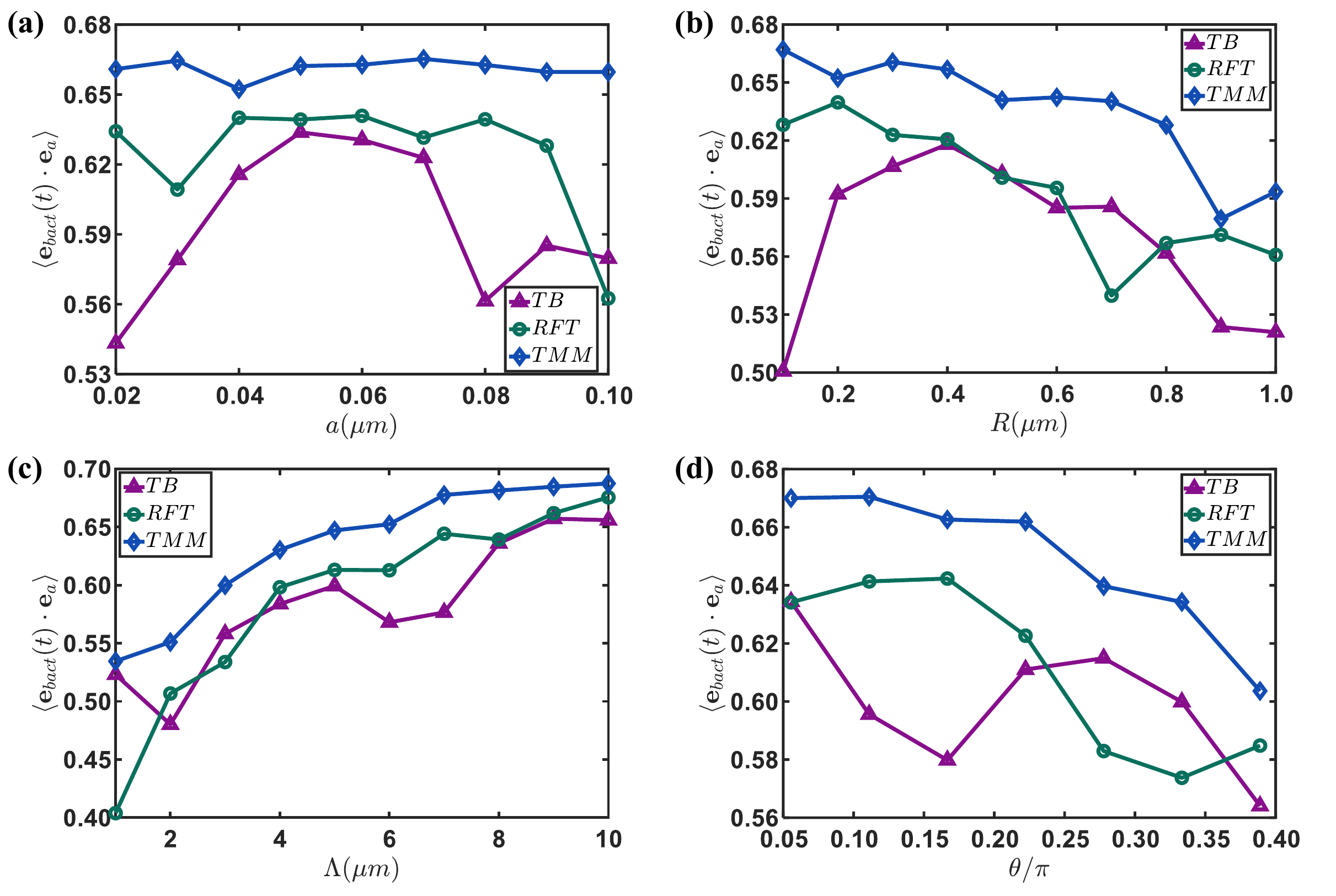}
\caption{Mean directional cosine of the bacterial forward direction as functions of (a) filament radius $a$, (b) helix radius $R$, (c) contour length $\Lambda$, and (d) pitch angle $\theta$, obtained from the chiral TB model, RFT, and TMM, respectively.}
\label{fig:fig9}
\end{figure}

As demonstrated in Fig.~\ref{fig:fig9}(a), the filament radius has a negligible effect on the mean directional cosine of the bacterial forward direction. The mean directional cosine shown in Fig.~\ref{fig:fig9}(b) exhibits a slight decrease with increasing helix radius. Fig.~\ref{fig:fig9}(c) indicates that increasing the contour length of the flagellum results in a more stable bacterial forward motion. In contrast, the mean directional cosine is relatively insensitive to variations in pitch angle, as illustrated in Fig.~\ref{fig:fig9}(d). In particular, the mean directional cosine obtained from the TMM consistently exceeds those obtained from both the chiral TB model and the RFT. This suggests that hydrodynamic interactions between the cell body and the flagellum contribute to maintaining a more stable forward direction for bacteria. Moreover, the close agreement between the simulation results from the chiral TB model and the RFT further validates the chiral TB model for simulating bacterial Brownian motion.

\section{Summary and Conclusions}
\label{sec4}
We characterize the symmetry and elongation of bacterial trajectories using the eigenvalues of the radius of gyration tensor. The directionality ratio quantifies the linearity of the trajectory, while the mean directional cosine of the bacterial forward direction assesses the stability of the forward motion. Numerical simulations indicate that increasing the contour length and helix radius of the flagellum leads to more elongated trajectories and enhances their linearity. Notably, the mean directional cosine primarily depends on the flagellar contour length, with longer lengths promoting more stable forward motion.

Importantly, our simulations demonstrate that the chiral TB model effectively simulates the Brownian motion of the bacteria. The agreement between the chiral TB model and the RFT simulations improves with increasing contour length. For contour lengths $\Lambda\ge 5.0$~\si{\mu m}, helix radii $R$ within $0.2\le R\le 0.5$~\si{\mu m}, pitch angles $\pi/6\le\theta\le2\pi/9$, the three simulation methods for bacterial Brownian motion show high consistency. In particular, the morphological parameters of \emph{Escherichia coli} fall precisely within this range~\cite{Darnton2007}. Comparison of the trajectories of the chiral TB model, RFT and TMM confirms the ability of the chiral TB model to capture key features of the Brownian bacterial motion, thus validating its effectiveness in simulating the Brownian bacterial motion.

The chiral TB model simplifies bacteria into two bodies but retains specific bacterial morphological characteristics while greatly reducing computational costs. This model is particularly effective for studying emergent behaviors in large bacterial populations, such as self-organization and active turbulence. Unlike many previous models, which often neglected detailed bacterial morphologies and hydrodynamic interactions due to computational cost, the chiral TB model effectively captures the flagellar morphology and chirality. Although the model simplifies the system by neglecting the hydrodynamic interactions among flagella, incorporating the hydrodynamic interactions between the spherical cell bodies, calculated using Rotne-Prager-Yamakawa tensors~\cite{Wajnryb2013,Vizsnyiczai2020} or TMM~\cite{Liu2025A}, can further improve the simulation accuracy.

\section*{Acknowledgements}
We gratefully acknowledge the computational resources and support provided by the Beijing Computational Science Research Center.
\appendix

\section{Hydrodynamic Resistance Matrices of a Chiral Two-body Model}
\label{Appendix}
A rotating helical flagellum along the $x$-axis, with helix radius $R$, pitch $\lambda$, axial length $L$, filament radius $a$, contour length $\Lambda=L/\cos\theta$ and pitch angle $\theta$, where $\tan\theta=2\pi R/\lambda$. The center of the flagellum is denoted as $\textbf{r}_{f}$. The chiral body model of the flagellum can be derived using RFT~\cite{Di2011,Dvoriashyna2021}. The hydrodynamic force and torque densities exerted by each filament element on the fluid are computed by RFT~\cite{Gray1955,Chwang1975,Johnson1979}:
\begin{equation}
\begin{split}
&\mathbf{f}=\mathbf{R}\cdot\mathbf{U},\\
&\mathbf{n}=(\mathbf{r}-\mathbf{r}_{f})\times\left(\mathbf{R}\cdot\mathbf{U}\right).
\label{eq:refnameA01}
\end{split}
\end{equation}
where $\mathbf{U}$ are the translational velocities of each segment. The centerline position of the left-handed helical flagellum is:
\begin{equation}
\begin{split}
&\mathbf{r}-\mathbf{r}_{f}=\left[x,R\sin\varphi,R\cos\varphi\right], x\in[-\frac{L}{2},\frac{L}{2}].
\label{eq:refnameA02}
\end{split}
\end{equation}
where $\varphi=\frac{2\pi}{\lambda}x+\varphi_{0}$ is the phase and $\varphi_{0}$ is the initial phase. The matrix $\mathbf{R}$ is the local hydrodynamic interaction matrix:
\begin{equation}
\begin{split}
&\mathbf{R}=k_{\parallel}\hat{\mathbf{t}}\otimes\hat{\mathbf{t}}+k_{\perp}\left(\mathbb{I}-\hat{\mathbf{t}}\otimes\hat{\mathbf{t}}\right).
\label{eq:refnameA03}
\end{split}
\end{equation}
where $\otimes$ is the Kronecker product, and $\hat{\mathbf{t}}$ is the local tangential unit vector:
\begin{equation}
\begin{split}
&\hat{\mathbf{t}}=\left(\cos\theta,\sin\theta\cos\varphi,-\sin\theta\sin\varphi\right).
\label{eq:refnameA04}
\end{split}
\end{equation}
The Gray and Hancock drag coefficients are\cite{Gray1955,Chwang1975}:
\begin{equation}
\begin{split}
&k_{\parallel}=\frac{2\pi\mu}{\ln(2\lambda/a)-1/2},\\
&k_{\perp}=\frac{4\pi\mu}{\ln(2\lambda/a)+1/2}.
\label{eq:refnameA05}
\end{split}
\end{equation}
where $\mu$ is the dynamic viscosity. The net force $\mathbf{F}_{f}$ and torque $\mathbf{T}_{f}$ exerted on the fluid by a rigid helical flagellum, with a translational velocity $\mathbf{U}_{f}$ and rotational velocity $\mathbf{W}_{f}$ satisfy the following relations:
\begin{equation}
\begin{split}
&\mathbf{F}_{f}=\mathbf{A}\cdot\mathbf{U}_{f}+\tilde{\mathbf{B}}\cdot\mathbf{W}_{f},\\
&\mathbf{T}_{f}=\mathbf{B}\cdot\mathbf{U}_{f}+\mathbf{C}\cdot\mathbf{W}_{f}.
\label{eq:refnameA06}
\end{split}
\end{equation}
The velocity at each point $\mathbf{r}$ on the rigid flagellum is given by:
\begin{equation}
\begin{split}
\mathbf{U}=\mathbf{U}_{f}+\mathbf{W}_{f}\times\left(\mathbf{r}-\mathbf{r}_{f}\right).
\label{eq:refnameA07}
\end{split}
\end{equation}
The net force $\mathbf{F}_{f}$ and torque $\mathbf{T}_{f}$ relative to the center of the flagellum are given by:
\begin{equation}
\begin{split}
&\mathbf{F}_{f}=\frac{1}{\cos\theta}\int_{-L/2}^{L/2}\mathbf{R}\cdot\mathbf{U}dx,\\
&\mathbf{T}_{f}=\frac{1}{\cos\theta}\int_{-L/2}^{L/2}\left(\mathbf{r}-\mathbf{r}_{f}\right)\times\left(\mathbf{R}\cdot\mathbf{U}\right)dx.
\label{eq:refnameA08}
\end{split}
\end{equation}
The resistance matrices of $\mathbf{A}$, $\mathbf{B}$ and $\mathbf{C}$ can be expressed as
\begin{equation}
\begin{split}
&\mathbf{A}=X_{\parallel}^{A}\mathbf{e}_{a}\otimes\mathbf{e}_{a}+X_{\perp}^{A}\left(\mathbb{I}-\mathbf{e}_{a}\otimes\mathbf{e}_{a}\right),\\
&\mathbf{B}=X_{\parallel}^{B}\mathbf{e}_{a}\otimes\mathbf{e}_{a}+X_{\perp}^{B}\left(\mathbb{I}-\mathbf{e}_{a}\otimes\mathbf{e}_{a}\right),\\
&\mathbf{C}=X_{\parallel}^{C}\mathbf{e}_{a}\otimes\mathbf{e}_{a}+X_{\perp}^{C}\left(\mathbb{I}-\mathbf{e}_{a}\otimes\mathbf{e}_{a}\right).
\label{eq:refnameA09}
\end{split}
\end{equation}
where $\mathbf{e}_{a}$ is the unit vector of the flagellar axis direction, and the matrix elements of the flagellar resistance matrix are given by:
\begin{equation}
\begin{split}
&X_{\parallel}^{A}=\Lambda\left[k_{\parallel}\cos^{2}\theta+k_{\perp}\sin^{2}\theta\right],\\
&X_{\perp}^{A}=\Lambda\left[k_{\parallel}\frac{\sin^{2}\theta}{2}+k_{\perp}\frac{1+\cos^{2}\theta}{2}\right],\\
&X_{\parallel}^{B}=RL\sin\theta\left(k_{\perp}-k_{\parallel}\right),\\
&X_{\perp}^{B}=-\frac{1}{2}RL\sin\theta\left(k_{\perp}-k_{\parallel}\right),\\
&X_{\parallel}^{C}=\Lambda R^{2}\left[k_{\parallel}\sin^{2}\theta+k_{\perp}\cos^{2}\theta\right],\\
&X_{\perp}^{C}=\Lambda\left[k_{\perp}(\frac{R^{2}}{2}+\frac{L^{2}}{12})+\left(k_{\parallel}-k_{\perp}\right)\sin^{2}\theta(\frac{R^{2}}{2\gamma^{2}}+\frac{L^{2}}{24})\right].
\label{eq:refnameA10}
\end{split}
\end{equation}
where $\gamma=\tan\theta=2\pi R/\lambda$. The resistance matrix of the flagellum is 
\begin{equation}
\begin{split}
\mathcal{R}_{t}=
\left(\begin{matrix} 
A_{f} & B_{f}^{T} \\
B_{f} & C_{f}
\end{matrix}\right).
\label{eq:refnameA11}
\end{split}
\end{equation}
The resistance matrix of a spherical cell body is
\begin{equation}
\mathcal{R}_{b}=\left(\begin{matrix}6\pi\mu R_{b}\mathbf{I}_{3} & \mathbf{0}_{3}\\ 
\mathbf{0}_{3} & 8\pi\mu R_{b}^{3}\mathbf{I}_{3} \end{matrix}\right).
\label{eq:refnameA12}
\end{equation}
where $\mathbf{I}_{3}$ is a $3\times3$ identity matrix, and $\mathbf{0}_{3}$ is a $3\times3$ zero matrix.

\bibliographystyle{unsrt} 
\bibliography{example}
\end{document}